\documentclass{moriond}

\usepackage{amsmath}

\def\dd{\mathrm{d}}

\def\be{\begin{equation}}
\def\ee{\end{equation}}
\def\bea{\begin{eqnarray}}
\def\eea{\end{eqnarray}}

\providecommand{\href}[2]{#2}

\newcommand{\pt}{p_{\text{T}}}

\newcommand{\pth}{{p_{\text{T,$H$}}}}

\newcommand{\noun}[1]{{\scshape #1}}

\newcommand{\POWHEG}{\noun{Powheg}}

\newcommand{\minlo}{{\noun{MiNLO$^{\prime}$}}}
\newcommand{\minnlo}{{\noun{MiNNLO$_{\rm PS}$}}}

\newcommand{\PYTHIA}[1]{\noun{Pythia{#1}}}

\usepackage{xcolor}

\def\ms{${\overline {\rm MS}}$}

\usepackage{etoolbox}
\makeatletter
\patchcmd{\@sect}{#8}{\boldmath #8}{}{}
\let\ori@chapter\@chapter
\def\@chapter[#1]#2{\ori@chapter[\boldmath#1]{\boldmath#2}}
\makeatother

\newcommand{\Photo}{\includegraphics[height=30mm]{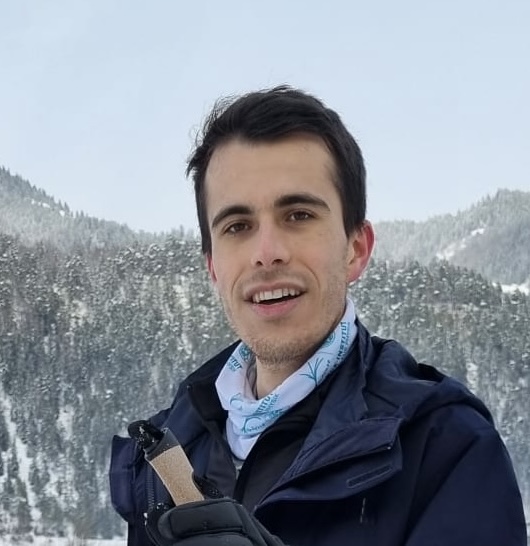}}

\begin{document}
\vspace*{4cm}
\title{NNLO+PS predictions for Higgs production via bottom annihilation}

\author{ Christian Biello }

\address{Max-Planck Institute for Physics,\\
Boltzmannstr. 8, 85748 Garching, Germany}

\maketitle\abstracts{
We report on the implementation of a new NNLO+PS event generator for the Higgs production via bottom annihilation\,\cite{our5FS} , using the  \minnlo\,  method in the \POWHEG\, framework. The calculation has been carried out in the five flavour scheme (5FS), where the bottom mass is neglected. We compare our results against fixed-order predictions at NNLO as well as resummed predictions at next-to-next-to-leading-logarithmic (NNLL) accuracy. We also present a preliminary study in the four flavour scheme (4FS) setup, achieving a new level of precision in the massive scheme.\\
\begin{flushright} MPP-2024-101 \end{flushright}
}

\section{Introduction}
Accurate simulation of $b\bar{b}H$ production at the LHC is crucial for identifying deviations from the Standard Model predictions and distinguishing between new physics signals and background noise. This mode directly probes the bottom Yukawa coupling and serves as the primary background for studying double Higgs production, enhancing the LHC's potential to uncover new phenomena and deepen our understanding of particle physics.\\

\noindent In particular, Higgs production via bottom fusion is theoretically intriguing for exploring different flavor schemes. In the five-flavor scheme (5FS), the bottom quark is treated as massless, allowing for the resummation of collinear logarithmic terms through DGLAP evolution. However, this approach is not able to capture power corrections related to the bottom mass, leading to less accurate descriptions of bottom kinematic distributions. Alternatively, the decoupling scheme treats bottoms as massive quarks within a four-active-flavor setup (4FS), explicitly considering mass effects but without resumming large logarithms. Total cross-sections are currently known from fixed-order calculations: N3LO QCD in 5FS\,\cite{duhr} and NLO QCD in 4FS\,\cite{dittmaier}.\\

\noindent Combining NNLO QCD calculations with parton showers (NNLO+PS) is a key challenge in collider theory. It's crucial for linking precise theory predictions with accurate experimental measurements. For $b\bar{b}H$ production, the first matching was performed for NLO predictions in 4FS using the MC@NLO\,\cite{MCatNLO} and \POWHEG\,\cite{NLOPWG} method. Here we focus on the first fully-differential calculation of NNLO QCD matched to parton shower using the \minnlo\, method.

\section{The method}
\noindent \POWHEG\,\cite{POWHEG} is a fully tested method for the matching of NLO predictions with parton shower which avoids the introduction of an unphysical matching scales. \POWHEG\, generates the hardest emission which alone gives the correct NLO result with an appropriate Sudakov form factor and interfaces the events with a Shower Monte Carlo for the subsequent radiations. The starting point of our calculation is the Higgs plus one jet production described by the \POWHEG\, formula,
\be
	\mathrm{d}\sigma_{\text{HJ}}=\mathrm{d}\Phi_{\text{HJ}} \, \bar B^{\text{\POWHEG}}\, \left\{ \Delta_{\text{pwg}}(\Lambda_{\text{pwg}})+\mathrm{d}\Phi_{\text{rad}} \Delta_{\text{pwg}}(p_{\text{T,rad}})\frac{R_{\text{HJ}}}{B_{\text{HJ}}} \right\},
\ee
where $\dd\Phi_{\text{HJ}}$ is the HJ phase space, $\Delta_{\text{pwg}}$ is the \POWHEG\, Sudakov form factor with a cutoff $\Lambda_{\text{pwg}}=0.89$ GeV. In addition, $\dd\Phi_{\text{rad}}$ and $p_{\text{T,rad}}$ are the phase space measure and the transverse momentum of the radiation with respect to HJ production.\\

\noindent The \minlo\, procedure\,\cite{MiNLO} improved NLO multijet calculations with the appropriate choice of scales and with the inclusion of a Sudakov form factor: this avoids an unphysical merging scale.  \minnlo\, empowers the \minlo\, master formula with other ingredients from the color singlet transverse momentum resummation. The $\bar B$ function thus reads:
\be
	\bar B^{\text{\minnlo}}=e^{-\tilde S(\pth)}\left\{B\left(1+\tilde S^{(1)}\right)+V+\int\dd \Phi_{\text{rad}}R + D^{(\geq 3)}(\pth) F^{\text{corr}}\right\},
\ee
where the Born $B$, virtual $V$ and real $R$ contributions are evaluated using $\pth$ as the renormalisation scale for the strong couplings. Here $\tilde S(\pth)$ is the \minlo\, Sudakov form factor and $\tilde S^{(1)}$ is its first order coefficient. The extension of \minnlo\, compared to \minlo\, is encoded in the $D^{(\geq 3)}$ terms multiplied by a spreading function $F^{\text{corr}}$. The $D^{(\geq 3)}$ terms contain all the required ingredients to obtain the NNLO accuracy for inclusive observables.

\section{Results in 5FS}
We provide numerical predictions for Higgs production via bottom-quark annihilation at the LHC, for 13 TeV center-of-mass energy. The mass of the Higgs boson is set to 125 GeV and its width to zero. Using the 5FS with massless bottom quarks, we employ a non-zero Yukawa coupling in \ms\, scheme, derived from $m_b(m_b)=4.18$ GeV and evolved via four-loop running. Our default setup uses the NNLO set of NNPDF40 parton densities with 5 active flavors and $\alpha_s(m_Z)=0.118$\,\cite{NNPDF}. Theory scale uncertainty is assessed with a standard 7-point scale variation. For predictions matched to a parton shower (PY8), we use \PYTHIA{8} with the A14 tune. We maintain Higgs boson stability and disable effects such as hadronization, multi-parton interactions (MPI), and QED radiation.\\

\begin{figure}
\begin{minipage}{0.49\linewidth}
\centerline{\includegraphics[width=1\linewidth]{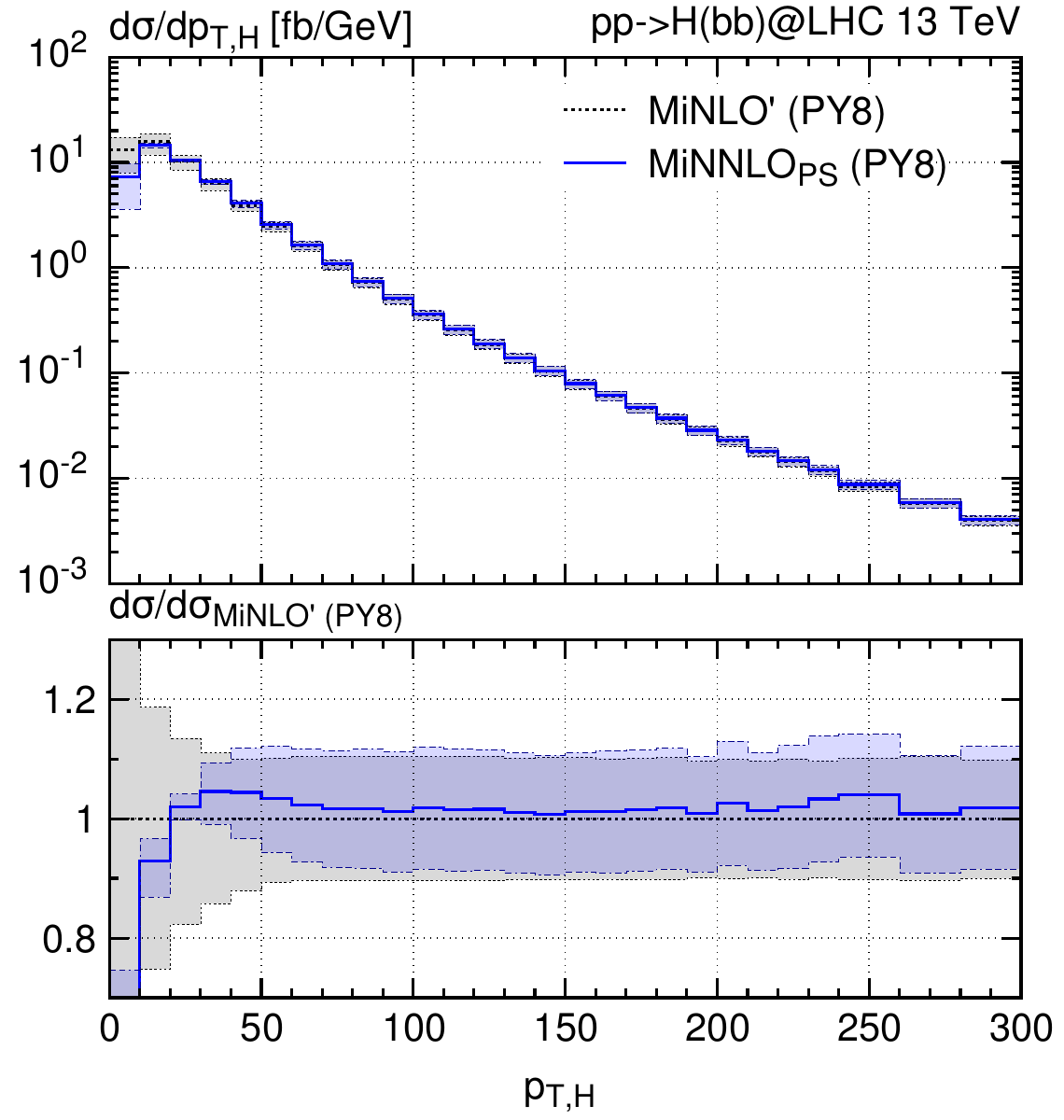}}
\caption[]{Higgs transverse spectrum from \minlo\, and \minnlo\, predictions with the Shower.}
\label{fig:minlopvsminnlo}
\end{minipage}
\hfill \hspace{0.3cm}
\begin{minipage}{0.49\linewidth}
\centerline{\includegraphics[width=1\linewidth]{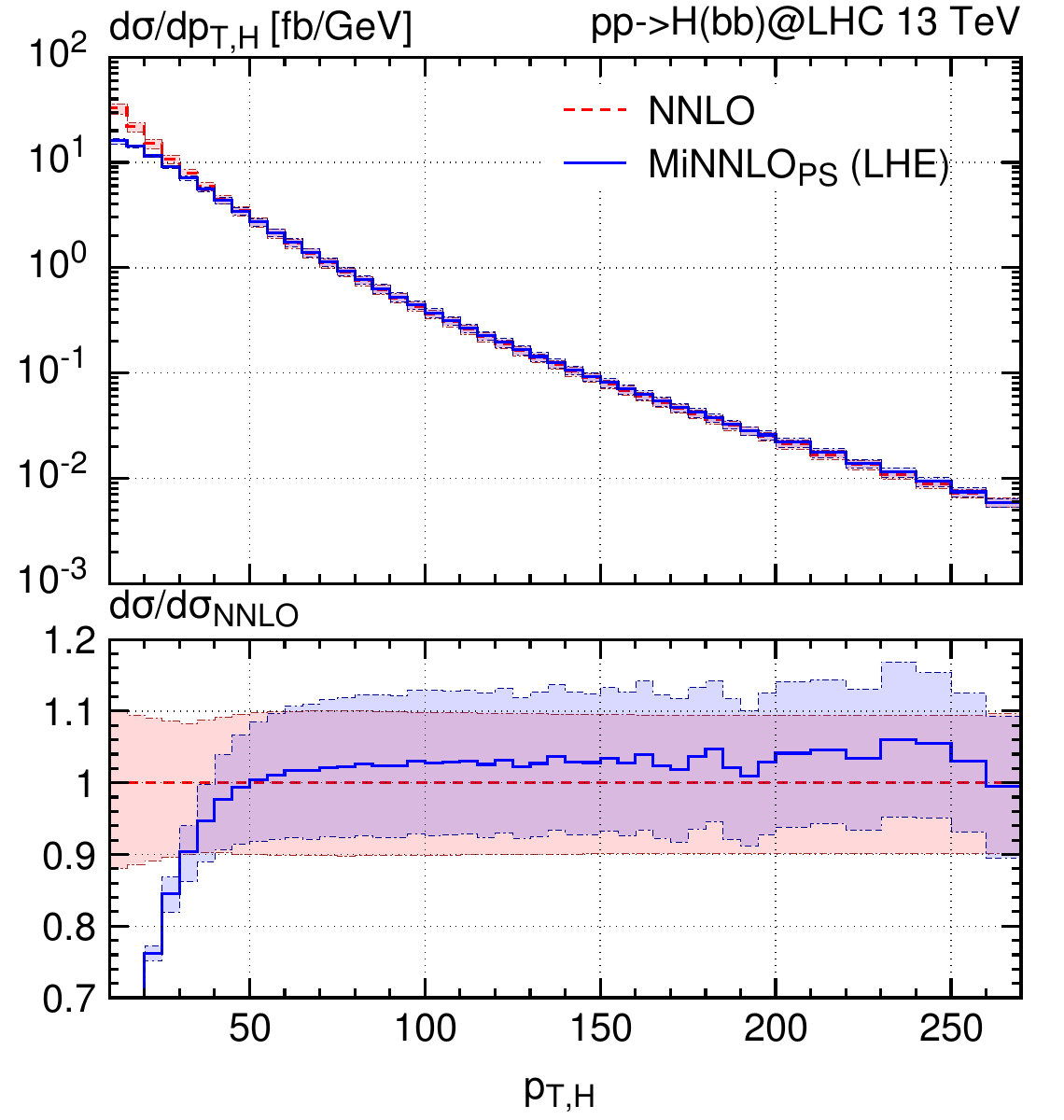}}
\caption[]{Comparison of the fixed-order NNLO prediction with \minnlo\, at LHE level.}
\label{fig:minnlovsNNLO}
\end{minipage}
\end{figure}

\noindent Using the \minnlo\, method, we obtain a total inclusive cross section of $0.509(8)_{-5.3\%}^{+2.9\%}$\,pb. Such value is in agreement with the fixed-order result of $0.518(2)_{-7.5\%}^{+7.2\%}$\,pb from {\sc SusHi}\,\cite{Sushi}. \minlo\, predicts a higher value of $0.571(1)_{-22.7\%}^{+17.4\%}$\,pb for the fully inclusive total cross-section, while it is compatible with the \minnlo\, result for high values of $\pth$ (see Figure \ref{fig:minlopvsminnlo}), where both generators effectively provide NLO accuracy. We validate our \minnlo\, generator against the fixed-order calculation\,\cite{NNLO}. From Figure \ref{fig:minnlovsNNLO}, it is clear that our predictions are agreement with the fixed-order predictions in the high  $p_{\text{T,H}}$ regime. However, the fixed-order calculation produces a divergence for vanishing $p_{\text{T,H}}$, while the \minnlo\, prediction remains finite.\\

\begin{figure}
\begin{minipage}{0.49\linewidth}
\centerline{\includegraphics[width=1\linewidth]{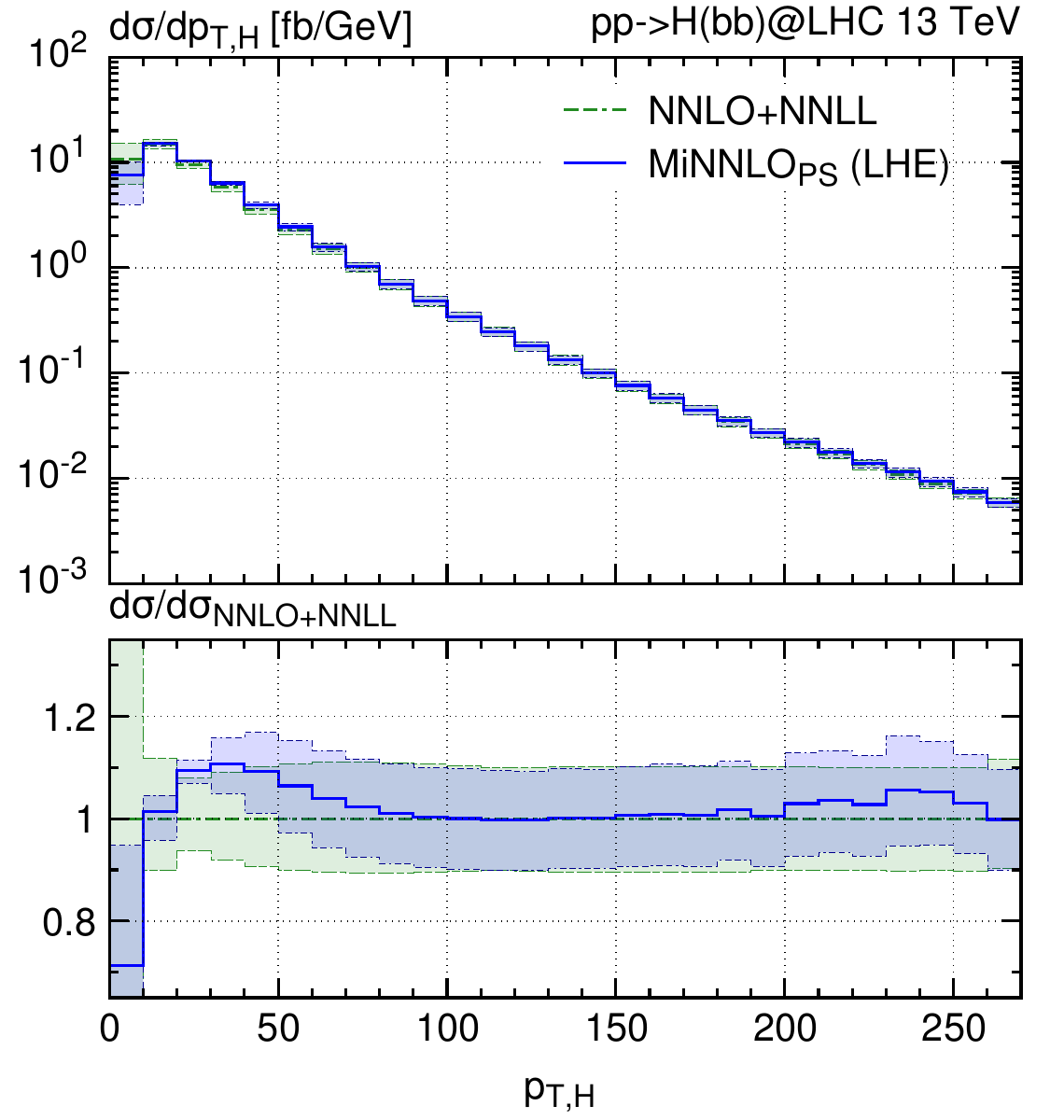}}
\end{minipage}
\hfill
\begin{minipage}{0.49\linewidth}
\centerline{\includegraphics[width=1\linewidth]{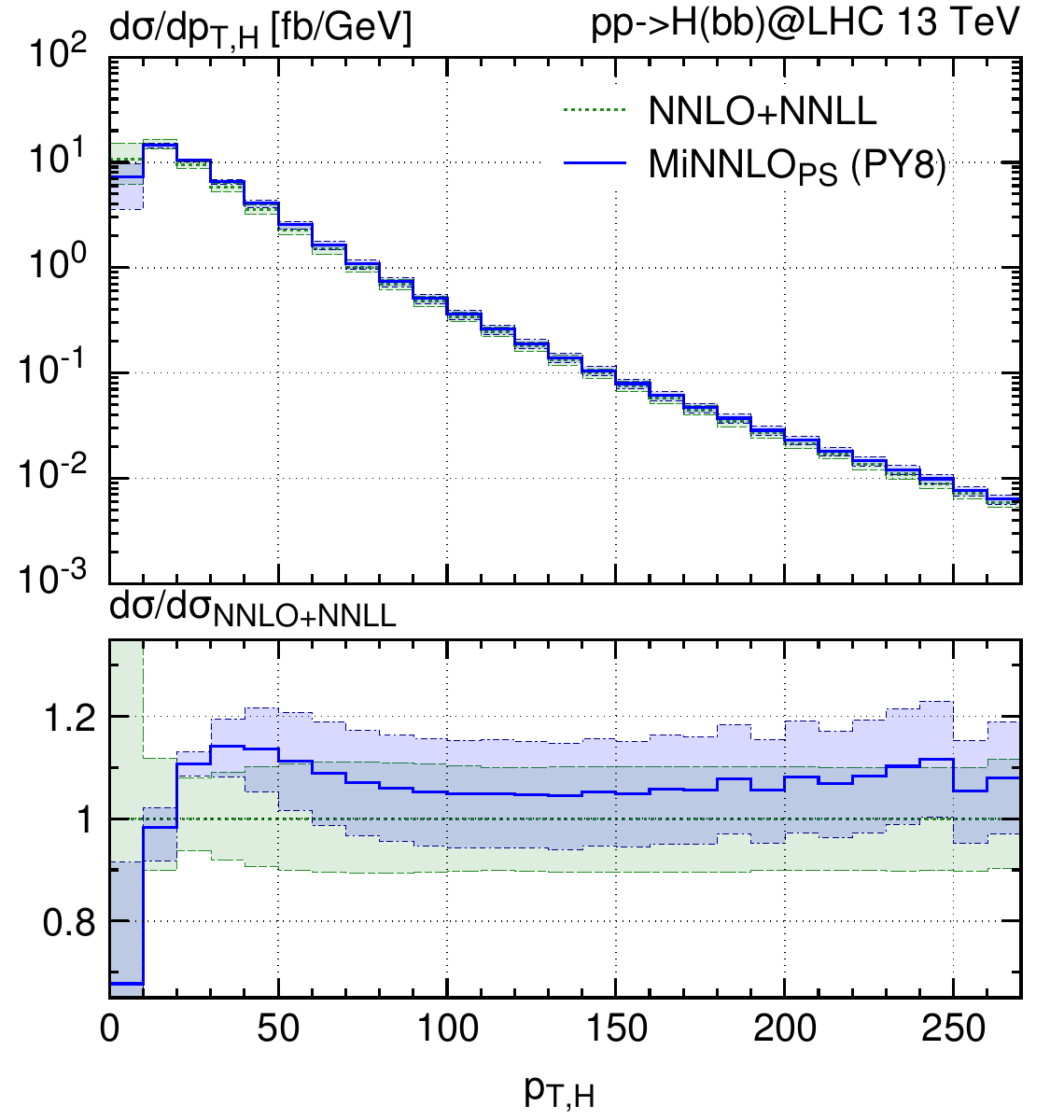}}
\end{minipage}
\caption[]{Predictions for the Higgs transverse spectrum from the analytic resummation at NNLO+NNLL compared with our \minnlo\, generator before (left) and after (right) the Parton Shower.}
\label{fig:NNLL}
\end{figure}

\noindent We further compare the \minnlo\, results with the analytic resummation at NNLO+NNLL\,\cite{NNLL} in Figure \ref{fig:NNLL}. Here we provide the \minnlo\, predictions at the LHE level in the left plot while the right one shows the results after showering with \PYTHIA{8}. The agreement is acceptable for small value of the transverse momentum. In the high $\pt$ region, the resummed results closely match with NNLO+NNLL before the shower. This agreement surpass the fixed-order comparison (Figure \ref{fig:minnlovsNNLO}), highlighting the impact of resummation. As observed in the second plot of Figure \ref{fig:NNLL}, the inclusion of the parton shower clearly affects high transverse momenta, leading to an approximate 10\% increase in the  \minnlo\, prediction. However, this modification remains contained within the scale-uncertainty bands.

\section{Preliminary studies in 4FS}
The 4FS implementation of the \minnlo\, generator\,\cite{our4FS} is more complex due to the higher multiplicity and the presence of massive colored final states. We present for the first time some preliminary results at the integrated level. We adapted the \minnlo\, method for heavy-quark pair production plus a color singlet for $b\bar{b}H$ production by taking into account the Yukawa scale dependence. The two-loop amplitude for massive bottom quarks is not known, but we can perform an approximation in order to capture the amplitude up to power corrections in the bottom mass. Starting from the amplitude with massless bottom quarks\,\cite{twoloop}, we apply the massification procedure\,\cite{massification}. We use a generalised approach which includes the leading mass effects of the bottom quarks in the loops\,\cite{massificationloops}.\\

\noindent By choosing the Higgs mass for the Yukawa scale and a quarter of the transverse invariant mass for the Born strong couplings, the NLO+PS generator gives a cross-section of $0.381(2)_{-16\%}^{+20\%}$\,pb. This value is  about $55\%$ smaller than the NLO 5FS prediction. The theoretical tension between the two schemes shows the importance of the collinear logarithmic contributions. Conversely, \minnlo\, predicts $0.464(9)_{-13\%}^{+14\%}$\,pb which is closer to the NNLO 5FS values. Indeed, for the first time, we are observing an agreement of the predictions in the two schemes within the scale uncertainty.

\section{Conclusion}
We presented the first matching of NNLO predictions with parton shower for the Higgs production via bottom fusion using the \minnlo\, method. We discussed the predictions in the 5FS and introduced the current studies in the 4FS with remarkable preliminary results.

\section*{Acknowledgments}

I am indebted to Chiara Savoini and Simone Zoia for  useful conversations and checks during the conference regarding the 4FS calculation. I express my gratitude to the organisers of the Rencontres de Moriond for giving me the opportunity to partecipate with a grant.

\newpage

%

\end{document}